\documentclass[aps, prl,10pt,notitlepage,nofootinbib,superscriptaddress,showkeys,twocolumn]{revtex4-1}
\usepackage{amstext,amsmath,amssymb,amsfonts,bbm, amsthm}
\usepackage[latin1]{inputenc}
\usepackage{fancyhdr}
\usepackage{graphicx}
\usepackage{hyperref}
\usepackage{tocvsec2}

\usepackage{color}

\def\beq{\begin{equation}}
\def\be{\begin{equation}}
\def\ee{\end{equation}}
\def\bes{\begin{eqnarray}}
\def\ees{\end{eqnarray}}

\def\f{\frac}

\def\pp{\partial}

\begin{document}
%%%%%%%%%%%%%%%%%%%%%%%%%%%%%%%%%%%%%%%%%%%%%%%%%%%

\title{\large \bf On the inertia of heat}

\author{{Matteo Smerlak}}\email{smerlak@aei.mpg.de}
\affiliation{Max-Planck-Institut f\"ur Gravitationsphysik, Am M\"uhlenberg 1, D-14476 Golm, Germany}

\date{\small\today}

%%%%%%%%%%%%%%%%%%%%%%%%%%%%%%%%%%%%%
\begin{abstract}\noindent
Does heat have inertia? This question is at the core of a long-standing controversy on Eckart's dissipative relativistic hydrodynamics. Here I show that the troublesome inertial term in Eckart's heat flux arises only if one insists on defining thermal diffusivity as a \emph{spacetime constant}. I argue that this is not the most natural definition, and that all confusion disappears if one considers instead the space-dependent \emph{comoving diffusivity}, in line with the fact that, in the presence of gravity, space is an \emph{inhomogeneous medium}.
\end{abstract}
%%%%%%%%%%%%%%%%%%%%%%%%%%%%%%%%%%%%%%
\maketitle

Eckart was the first to consider the problem of heat conduction in general relativity. In the seminal 1940 paper \cite{Eckart1940}, he proposed as that the general-relativistic the heat equation
\be\label{eckartheat}
u^{a}\nabla_{a}T+\theta T+\nabla_{a}q^{a}=0.
\ee
Here $u^{a}$ is the $4$-velocity of the thermal medium, $\theta=\nabla_{a}u^{a}$ its expansion, and $q^{a}$ the heat flux.\footnote{Equivalently, this equation can be written as the continuity equation $\nabla_{a}\rho^{a}=0$ for the energy flux $\rho^{a}=Tu^{a}+q^{a}$ (in units where the heat capacity is $1$).} To close this equation, he proposed the constitutive relation
\be\label{ansatz}
q_{b}=-\kappa_{0}(D_{b}T+a_{b}T)
\ee
where $D_{b}T=(u^{a}u_{b}+\delta^{a}_{b})\nabla_{a}T$ is the spatial gradient of $T$ (in the local rest frame associated to $u^{a}$), $a^{b}=u^{a}\nabla_{a}u^{b}$ the local acceleration, and $\kappa_{0}$ the thermal diffusivity. This ansatz, he argued, is the simplest one that is consistent with the second law of hydrodynamics. 

In spite of early criticisms of Eckart's approach \cite{PhamMauQuan1955,Bennoun1965}, the system \eqref{eckartheat}-\eqref{ansatz} (or a variant thereof due to Landau and Lifshitz \cite{Landau1987}) was considered the standard formulation of relativistic thermal dynamics until the late 70's. Together with the relativistic Navier-Stokes equations, it formed the theory of ``dissipative relativistic hydrodynamics''; its connection to relativistic kinetic theory was believed to be well understood \cite{Ehlers1974}. 

The situation changed drastically with two major contributions in the field: Israel and Stewart's development of a ``second-order'' formalism bridging between the kinetic (microscopic) and hydrodynamic (macroscopic) levels of description of relativistic transport phenomena \cite{Israel1976a,Israel1979}; and the discovery by Hiscock and Lindblom of ``generic instabilities'' in Eckart's ``first-order'' formalism \cite{Hiscock1985,Hiscock1985a}. (In this context, the expressions ``first order'' and ``second order'' refer to the parabolic and hyperbolic nature of the corresponding system of partial differential equations.) The consensus soon became that Eckart's theory is both \emph{acausal} and \emph{unstable}, and thus ``unacceptable''. Given the high complexity of the Israel-Stewart ``extended thermodynamics'', with involves many more dynamical fields than the standard hydrodynamic ones (temperature, velocity, pressure), the field has remained ---to this day!--- without an accepted theory of relativistic dissipative phenomena. This is all the more frustrating as new, important applications have arisen in high-energy physics, in particular regarding the dynamics of the quark-gluon plasma.

%\be
%\nabla_{a}\Big(T u^{a}u^{b}+p(g^{ab}+u^{a}u^{b})+2u^{(a}q^{b)}+\pi^{ab}\Big)=0,
%\ee
%Here $p$ is the pressure of the fluid and $\pi_{ab}$ the shear viscosity, for which Eckart also gave an expression in terms of local gradients. (We work in units where the heat capacity is $1$.)

The bottom line of the recent criticisms to Eckart's theory is the presence of a temperature-acceleration coupling $a_{b}T$ in the constitutive relation. This term, coined the ``inertia of heat'', has no analogue in Fourier's constitutive relation, and is allegedly responsible for the unphysical behavior of the heat equation \eqref{eckartheat}. Among the old and recent arguments to this effect, one reads that: it is not consistent with the relativistic version of the first principle \cite{Bennoun1965}; it depends on an arbitrary choice of frame \cite{Kampen1987}; it is not a gradient, while linear irreversible thermodynamics requires that all dissipative fluxes are gradients \cite{Sandoval-Villalbazo2009}; heat is a non-mechanical form of energy, so it should not have ``inertia'' \cite{Garcia-Colin2006,Garcia-Colin2007,Muschik2007}; it is not consistent with relativistic kinetic theory \cite{Sandoval-Villalbazo2009}; it is the source of the Hiscock-Lindblom ``generic instabilities'' \cite{Garcia-Perciante2009,Andersson2010}. 

My purpose in this note is not to address these objections head-on.\footnote{The issue of the consistency of Eckart's ansatz with relativistic kinetic theory will be addressed elsewhere \cite{Smerlaka}.} Rather, I wish to discuss the \emph{physical} origin of the inertia of heat. But before that, let me make three preliminary observations.

First, the issue of causality is not an issue at all. The usual Fourier heat equation also propagates signals faster than light, in patent contradiction with the much slower values of microscopic thermal velocities. All this means is that we should not forget that dissipative hydrodynamics is a long-time, large-scale approximation of kinetic theory, which breaks down when spacetime gradients arise on scales comparable to the mean free flight and mean free path. In its regime of validity, the heat equation is perfectly consistent with causality, and the same holds for Eckart's equation. This point was made repeatedly in the literature \cite{Weymann1967,Kampen1987}, and should by now be widely acknowledged. Furthermore, it has been showed that alternative ``hyperbolic theories'' of the Israel-Stewart type, which involve more dynamical variables and free functions, generally have the same physical content as parabolic theories \`a la Eckart \cite{Geroch1995}.

Second, the ``generic instabilities'' discovered by Hiscock and Lindblom \cite{Hiscock1985,Hiscock1985a} and referred to in \cite{Garcia-Perciante2009,Andersson2010} pertain to the coupled heat-fluid dynamics, i.e. the relativistic Fourier-Navier-Stokes system. The issue at stake here is whether it is consistent to include Eckart's heat flux \eqref{ansatz} into the stress-energy-momentum tensor $T_{ab}$ entering the hydrodynamical equation $\nabla_{a}T^{ab}=0$ for a dissipative fluid; some authors have argued that it should not \cite{Garcia-Colin2006}. Here, I would like to focus instead on the heat dynamics itself, assuming a prescribed background fluid velocity $u^{a}$. This is as sensible approximation as Fourier's theory for heat conduction, and is valid in a regime where the backreaction of the heat flux on the medium's dynamics is negligible. In this approximation, the no-go result of \cite{Hiscock1985,Hiscock1985a}, however important in itself, is not relevant; in particular it must not prevent us from considering the issue of the inertia of heat in itself.

Third, there \emph{is} an easy way to understand the meaning of the inertia of heat: by considering \emph{thermal equilibrium} in static spacetimes. Denoting $\xi^{a}$ a timelike Killing vector and $\chi=(-\xi^{a}\xi_{a})^{1/2}$ the redshift factor, and observing that the acceleration of the hydrostatic congruence $u^{a}=\xi^{a}/\chi$ is $a_{b}=D_{b}\chi/\chi,$ we see that $q_{b}=0$ is equivalent to $\chi T=\textrm{const}$, as prescribed by Tolman's law \cite{Tolman1930}. In other words, the inertial term in Eckart's ansatz is exactly what it takes to ensure that the equilibrium temperature distribution compensates gravitational redshifts. This argument is very compelling; in fact, it was almost always cited in the early reviews of relativistic hydrodynamics, as e.g. in \cite{Ehlers1974}.

%are not meaningful. As showed in \cite{Kostadt2000}, they arise only for \emph{ill-posed} Cauchy problems obtained by writing the heat equation \eqref{eckartheat} in a frame in which $u^{a}$ has non-zero spatial components. But setting initial data for $T$ on a hypersurface which is not comoving with $u^{a}$ explicitely violates the conditions of applicability of the hydrodynamic approximation: it means that ... One should not forget that the hydrodynamic limit of kinetic theory breaks the Lorentz symmetry, for that is a fact.

I now come to the thesis of this note: \emph{the disturbing inertial term in Eckart's ansatz comes from an awkward interpretation of the notion of ``thermal diffusivity''}. Let me explain.

In Eckart's approach, and in all subsequent work I am aware of, thermal diffusivity is defined as a \emph{constant} coefficient $\kappa_0$, as in \eqref{ansatz}. However, dimensional analysis teaches us that diffusivity has dimensions $(\textrm{length})^{2}/\textrm{time}$, and this implies that its value should change if the unit of time changes. Now, in a curved spacetime, this is precisely what happens \emph{from point to point}.

For clarity, assume as before that the thermal medium is in hydrostatic equilibrium, so that its instantaneous rest frame can be described by static \emph{comoving coordinates} $(t,x^{i})$ such that\footnote{We assume that $u^{a}$ is irrotational.}
\be
ds^{2}=-\chi(x)^{2}dt^{2}+q_{ij}dx^{i}dx^{j}.
\ee
Here $\chi$ is again the redshift factor and $q_{ij}$ is the comoving spatial metric, with associated covariant derivative $D_{i}$. In these coordinates, Eckart's heat equation become
\be
\chi^{-1}\pp_{t}T=\kappa_{0}(D_{i}+a_{i})(D^{i}T+a^{i}),
\ee
viz.
\be\label{comovingheat}
\pp_{t}T=\kappa_{0}\chi(\Delta T+2a^{i}D_{i}T+a^{2}T),
\ee
where $\Delta=D_{i}D^{i}$ is the spatial Laplacian. Now, defining the function
\be
\kappa(x)=\kappa_{0}\chi(x)
\ee
it is easy to see that \eqref{comovingheat} is equivalent to
\be\label{easyeckart}
\pp_{t}T=\Delta(\kappa T).
\ee
This is an interesting observation: Eckart's equation neatly simplifies when we use the function $\kappa(x)$ instead of the constant $\kappa_{0}$. In particular, all terms involving the acceleration $a^{i}$ disappear. But what is the interpretation of $\kappa(x)$?

My answer is that $\kappa(x)$ should be interpreted as the \emph{comoving diffusivity}, and heat conduction in general relativity as \emph{diffusion in a inhomogeneous medium}. 

The most useful definition of diffusivity is in terms of the mean square displacement of a Brownian particle whose probability density satisfies the corresponding heat equation. If $\langle d^{2}(x_{s},x_{0})\rangle$ is the mean of the squared spatial distance between the position $x_{s}$ of a Brownian particle at time $s$ and its starting point $x_{0}$, we can define the diffusivity as $\lim_{s\rightarrow0}\langle d^{2}(x_{s},x_{0})\rangle/6s$. Now, from \eqref{easyeckart}, we find that if $s$ is the proper time $\tau$ along a static worldline initiated at $x_{0}$, then
\be
\lim_{\tau\rightarrow0}\f{\langle d^{2}(x_{\tau},x_{0})\rangle}{6\tau}=\kappa_{0}, 
\ee
while if $s$ is the comoving time coordinate $t$, we have
\be
\lim_{t\rightarrow0}\f{\langle d^{2}(x_{t},x_{0})\rangle}{6t}=\kappa(x_{0}). 
\ee
This observation justifies the names \emph{proper diffusivity} for $\kappa_{0}$ and \emph{comoving diffusivity} for $\kappa(x)$. The point is that, while the former is constant in space, the latter is not (unless there is no gravity). This property is precisely what makes the comoving diffusivity physically relevant: it expresses the fact that, if time runs faster in some region of space, then diffusion is also faster in these regions. This is the origin of the ``inertia of heat''.

Of course, the notion of proper diffusivity is also useful, since it is the one related to the proper time measured by physical clocks. The point is that this \emph{invariant} notion is at variance with an intuition based on the notion of \emph{space} (as opposed to spacetime). Indeed, this notion of involves a foliation of spacetime indexed by the comoving time coordinate $t$, and therefore requires that diffusivity be measured in units of \emph{this} global time -- where it is not homogeneous but space-dependent. Thus, I propose that the unease with Eckart's inertial term comes from a clash between spatial intuitions and the covariant formulation of the heat equation \eqref{eckartheat}.

Let me now step back and reconsider the derivation of the general-relativistic heat equation from this perspective. In comoving coordinates, the thermal medium is inhomogeneous, with a space-dependent diffusivity $\kappa(x)$. As usual, we may write the continuity equation in the form
\be
\pp_{t}T=-D_{i}Q^{i}
\ee
with $Q^{i}$ the comoving heat flux, involving the gradient of $\kappa$ and $T$. Now, given that diffusivity is now a function of space, the form that $Q^{i}$ should take is not obvious at priori: should it take the ``Fourier'' form
\be
Q^{i}=-\kappa D^{i}T
\ee
or the ``Fokker-Planck'' form
\be\label{fp}
Q^{i}=-D^{i}(\kappa T),
\ee
or some other, hybrid form? This question is well-known in the literature on diffusion in inhomogeneous media \cite{VanKampen1988,Bringuier2011}. The accepted answer is that, unless there is a ``microscopic bias'' implying that the mean position $\langle x_{s}\rangle$ of Brownian particles changes with time, the correct form is the Fokker-Planck one \cite{Bringuier2011}. If we now remember that heat is carried by the very same Brownian particles which make up the thermal medium, and that in the comoving frame their mean velocity vanishes, we see that the right pick  for the heat equation is \eqref{fp}. This gives \eqref{easyeckart}, which we saw is indeed Eckart's equation (in hydrostatic equilibrium). Observe that \eqref{fp} is a gradient, as required by linear irreversible thermodynamics.

With hindsight, it is not particularly surprising that the relativistic heat equation is more easily interpreted in the comoving frame than in covariant form \eqref{eckartheat}, in spite of the fact that the corresponding time coordinate $t$ is not observable: dissipative phenomena  are \emph{by nature} alien to covariance. They are associated to the production of entropy, viz. with a thermodynamical \emph{arrow of time}. Now, it so happens that the natural time variable associated to this arrow is not the local proper time $\tau$: it is the global comoving time $t$.

\bibliographystyle{utcaps}
\bibliography{library}

\end{document}